\documentclass[a4paper]{article}

\usepackage[english]{babel}
\usepackage[utf8x]{inputenc}
\usepackage{amsmath}
\usepackage{graphicx}
\usepackage[colorinlistoftodos]{todonotes}
\usepackage{listings}
\usepackage{amssymb}
\usepackage{multicol}
\usepackage{algpseudocode}
\usepackage[normalem]{ulem}
\title{On Generating Permutations Under User-Defined Constraints}
\author{Dhruvil Badani\footnote{Grade 12, RN Podar School, Mumbai 400054, Maharashtra, India.}\\ dhruvilbadani@gmail.com}

\begin{document}
\maketitle

\begin{abstract}
In this paper, a method to generate permutations of a string under a set of constraints decided by the user is presented. The required permutations are generated without generating all the permutations.
\end{abstract}

\section{Introduction}

A string is traditionally a sequence of characters. A permutation of a string is a rearrangement of the characters S into a one-one correspondence with S itself.  A Matrix is a rectangular array of numbers, symbols, or expressions, arranged in rows and columns. The determinant is a value associated with a square matrix.

Permutations are used in error detection and correction algorithms. They are also used in fields of science such as combinatorial biology.  

Presented in this paper is a method to generate permutations under user-defined constraints. The method generates only those permutations that obey the condition, as opposed to generating all permutations and then checking if they satisfy the user-defined constraints. The permutations are generated in lexicographic order. Repetitive permutations are eliminated.

Section 2 mentions the problem. Section 3 shows how to represent the constraints. Section 4 provides an illustration of the algorithm, taking a sample string and a sample set of constraints. Section 5 describes the exact procedure to generate the permutations under the given set of constraints. Section 6 provides a conclusion to this paper. Finally, a set of references is mentioned.

\section{Problem Statement}
Generating the permutations of a string S (may contain duplicate characters) of length n such that it satisfies all constraints entered by the user.
Constraint refers to fixation of character(s) at position(s) by allowing or forbidding character(s) at the position(s). For example, if S is "abc1a23", the constraints \emph{can include but do not restrict themselves to one or many of the following}:
\\
\begin{description}
  \item[$\bullet$] The letter at position 2 should be a 
  \item[$\bullet$] The letter at position 2 should be a or b or 1
  \item[$\bullet$] The letter at position 3 should not be b
  \item[$\bullet$] The letter at position 3 should not be b or c
  \item[$\bullet$] The letter at position 1 should be a and the letter at    position 3 should be c
  \item[$\bullet$] The letter at position 1 should be a or c and the letter at position 4 should be b or 1
  \item[$\bullet$] The letter at position 1 should be a or 1 and the letter at position 5 should not be b or c
  
and many likewise constraints.
\end{description}

\section{Representation of Constraints}
Construct two $N\times N$ matrices. One matrix is used to store the permitted characters and the other is used to store the forbidden characters. 

Row i in each of the two matrices will store the permitted and forbidden characters at position i respectively, using its columns. 

For example, if S is "abcdae" and the set of constraints is:

\begin{description} 
  \item[$\bullet$] first letter should be a or c
  \item[$\bullet$] third letter should not be d or e
  \item[$\bullet$] fourth letter should be b
\end{description}

Then the "permitted matrix" will be of the form:
\\
\begin{multicols}{2}
$$\begin{bmatrix}
   a & c &   &   &   &   \\
     &   &   &   &   &   \\
     &   &   &   &   &   \\
   b &   &   &   &   &   \\
     &   &   &   &   &   \\
     &   &   &   &   &   \\
  \end{bmatrix}$$
The "forbidden matrix" will be of the form:
$$\begin{bmatrix}
     &   &   &   &   &   \\
     &   &   &   &   &   \\
   d & e &   &   &   &   \\
     &   &   &   &   &   \\
     &   &   &   &   &   \\
     &   &   &   &   &   \\
  \end{bmatrix}$$
\end{multicols}
Now, in order to make the implementation easier, a certain set of manipulations on the "permitted" and "forbidden" matrices are carried out. 
\\
\\
If a particular row i is empty in both- the permitted and forbidden matrices - it is implied that all letters are permitted at position i. Hence, in this case, the "permitted matrix" would look like this:
\\
$$\begin{bmatrix}
   a & c &   &   &   &   \\
   a & b & c & d & e &   \\
     &   &   &   &   &   \\
   b &   &   &   &   &   \\
   a & b & c & d & e &   \\
   a & b & c & d & e &   \\
  \end{bmatrix}$$
\\
\\
Note that the constraint on any particular position cannot mention to allow certain character(s) and forbid certain character(s) at that particular position. This is because:
\begin{description} 
  \item[$\bullet$] The letters which are neither permitted nor forbidden would create ambiguity.
  \item[$\bullet$] Allowing or forbidding certain character(s) implies allowing or forbidding those character(s) only.
\end{description}

Hence, a particular row i can be filled in either the permitted or the forbidden matrix - not both.
\\
\\
Keeping in mind the above facts, the "permitted matrix" is changed so that while implementing, we have to work only with the "permitted matrix" and not the "forbidden matrix." 
\\
\\
For example, in this case, the constraint on letter 3 is to forbid the characters d and e. This implies that the letters at position three can be a,b or c. Hence, the "permitted matrix" now looks like this:
\\
\\
\\
$$\begin{bmatrix}
   a & c &   &   &   &   \\
   a & b & c & d & e &   \\
   a & b & c &   &   &   \\
   b &   &   &   &   &   \\
   a & b & c & d & e &   \\
   a & b & c & d & e &   \\
  \end{bmatrix}$$
\\
\\
Hence, for applying the constraints now, we only need the "permitted matrix."
\\
\\
\begin{multicols}{2}
\section{Illustration}
$S="abacb"$
\\
$\therefore n=5 $
\\
Sort S in ascending order.
\\
$\therefore S="aabbc" $
\\
Construct the $n\times n$ $ie$ $5\times 5$ matrix A:
\\
$$\begin{bmatrix}
   a & a & b & b & c\\
   a & a & b & b & c\\
   a & a & b & b & c\\
   a & a & b & b & c\\
   a & a & b & b & c\\
  \end{bmatrix}$$
\\
\\
Set of constraints:
\begin{description} 
  \item[$\bullet$] First letter is a or b 
  \item[$\bullet$] Third letter is not c
\end{description}
$$\therefore "permitted"=
\begin{bmatrix}
   a & b &   &   & \\
     &   &   &   & \\
     &   &   &   & \\
     &   &   &   & \\
     &   &   &   & \\
  \end{bmatrix}$$
\\
\\
\\
\\
$$\therefore "forbidden"=
\begin{bmatrix}
     &   &   &   & \\
     &   &   &   & \\
   c &   &   &   & \\
     &   &   &   & \\
     &   &   &   & \\
  \end{bmatrix}$$
\\
\\
\\
Rows 2,4 and 5 are empty in both the "permitted matrix" and the "forbidden matrix." 
\\
$$\therefore "permitted"=
\begin{bmatrix}
   a & b &   &   & \\
   a & b & c &   &  \\
     &   &   &   & \\
   a & b & c &   &  \\
   a & b & c &   &  \\
  \end{bmatrix}$$
\\
Also, since letter c is not permitted at position 3, it is implied that the letters a and b are permitted.
\\
$$\therefore "permitted"=
\begin{bmatrix}
   a & b &   &   & \\
   a & b & c &   &  \\
   a & b &   &   & \\
   a & b & c &   &  \\
   a & b & c &   &  \\
  \end{bmatrix}$$
\\
Now, we need to work only with the "permitted matrix."\\
The computation is:
$$f\left(\begin{bmatrix}
   a & a & b & b & c\\
   a & a & b & b & c\\
   a & a & b & b & c\\
   a & a & b & b & c\\
   a & a & b & b & c\\
  \end{bmatrix},1\right)$$
\\
Since the level is 1, we carry out the computation of only those letters that are present in row 1 of the "permitted matrix" ie a and b. Hence, the computation is:
\\
$$a+f\left(\begin{bmatrix}
         a & b & b & c\\
         a & b & b & c\\
         a & b & b & c\\
         a & b & b & c\\
         \end{bmatrix},2\right),$$ \\ $$b+f\left(\begin{bmatrix}
         a & a & b & c\\
         a & a & b & c\\
         a & a & b & c\\
         a & a & b & c\\
         \end{bmatrix},2\right)$$ 
Notice that \\
$$a+f\left(\begin{bmatrix}
         a & b & b & c\\
         a & b & b & c\\
         a & b & b & c\\
         a & b & b & c\\
         \end{bmatrix},2\right),$$ \\ is present only once. This is in accordance with the procedure and is done to prevent repetitive permutations from being generated. The same applies for \\ $$b+f\left(\begin{bmatrix}
         a & a & b & c\\
         a & a & b & c\\
         a & a & b & c\\
         a & a & b & c\\
         \end{bmatrix},2\right)$$\\ Since the level now is 2, we carry out the computation for only those characters that are present in row 2 of the "permitted matrix" ie a,b and c. Keeping in mind the prevention of the generation of repetitive permutations, the computation is:\\
$$a+a+f\left(\begin{bmatrix}
         b & b & c\\
         b & b & c\\
         b & b & c\\
         \end{bmatrix},3\right),$$\\
$$a+b+f\left(\begin{bmatrix}
         a & b & c\\
         a & b & c\\
         a & b & c\\
         \end{bmatrix},3\right),$$\\
$$a+c+f\left(\begin{bmatrix}
         a & b & b\\
         a & b & b\\
         a & b & b\\
         \end{bmatrix},3\right),$$\\
$$b+a+f\left(\begin{bmatrix}
         a & b & c\\
         a & b & c\\
         a & b & c\\
         \end{bmatrix},3\right),$$\\
$$b+b+f\left(\begin{bmatrix}
         a & a & c\\
         a & a & c\\
         a & a & c\\
         \end{bmatrix},3\right),$$\\
$$b+c+f\left(\begin{bmatrix}
         a & a & b\\
         a & a & b\\
         a & a & b\\
         \end{bmatrix},3\right)$$\\  Since the level now is 3, we carry out the computation for only those characters that are present in row 3 of the "permitted matrix" ie a and b. Keeping in mind the prevention of the generation of repetitive permutations, the computation is:\\
$$a+a+b+f\left(\begin{bmatrix}
         b & c\\
         b & c\\
         \end{bmatrix},4\right),$$\\
$$a+b+a+f\left(\begin{bmatrix}
         b & c\\
         b & c\\
         \end{bmatrix},4\right),$$\\
$$a+b+b+f\left(\begin{bmatrix}
         a & c\\
         a & c\\
         \end{bmatrix},4\right),$$\\
$$a+c+a+f\left(\begin{bmatrix}
         b & b\\
         b & b\\
         \end{bmatrix},4\right),$$\\
$$a+c+b+f\left(\begin{bmatrix}
         a & b\\
         a & b\\
         \end{bmatrix},4\right),$$\\
$$b+a+a+f\left(\begin{bmatrix}
         b & c\\
         b & c\\
         \end{bmatrix},4\right),$$\\
$$b+a+b+f\left(\begin{bmatrix}
         a & c\\
         a & c\\
         \end{bmatrix},4\right),$$\\
$$b+b+a+f\left(\begin{bmatrix}
         a & c\\
         a & c\\
         \end{bmatrix},4\right),$$\\
$$b+c+a+f\left(\begin{bmatrix}
         a & b\\
         a & b\\
         \end{bmatrix},4\right),$$\\
$$b+c+b+f\left(\begin{bmatrix}
         a & a\\
         a & a\\
         \end{bmatrix},4\right)$$\\  Since the level now is 4, we carry out the computation for only those characters that are present in row 4 of the "permitted matrix" ie a,b and c. Keeping in mind the prevention of the generation of repetitive permutations, the computation is:\\
$$a+a+b+b+f\left(\begin{bmatrix}
         c\\
         \end{bmatrix},5\right),$$\\
$$a+a+b+c+f\left(\begin{bmatrix}
         b\\
         \end{bmatrix},5\right),$$\\
$$a+b+a+b+f\left(\begin{bmatrix}
         c\\
         \end{bmatrix},5\right),$$\\
$$a+b+a+c+f\left(\begin{bmatrix}
         b\\
         \end{bmatrix},5\right),$$\\
$$a+b+b+a+f\left(\begin{bmatrix}
         c\\
         \end{bmatrix},5\right),$$\\
$$a+b+b+c+f\left(\begin{bmatrix}
         a\\
         \end{bmatrix},5\right),$$\\
$$a+c+a+b+f\left(\begin{bmatrix}
         b\\
         \end{bmatrix},5\right),$$\\
$$a+c+b+a+f\left(\begin{bmatrix}
         b\\
         \end{bmatrix},5\right),$$\\
$$a+c+b+b+f\left(\begin{bmatrix}
         a\\
         \end{bmatrix},5\right),$$\\
$$b+a+a+b+f\left(\begin{bmatrix}
         c\\
         \end{bmatrix},5\right),$$\\
$$b+a+a+c+f\left(\begin{bmatrix}
         b\\
         \end{bmatrix},5\right),$$\\
$$b+a+b+a+f\left(\begin{bmatrix}
         c\\
         \end{bmatrix},5\right),$$\\
$$b+a+b+c+f\left(\begin{bmatrix}
         a\\
         \end{bmatrix},5\right),$$\\
$$b+b+a+a+f\left(\begin{bmatrix}
         c\\
         \end{bmatrix},5\right),$$\\
$$b+b+a+c+f\left(\begin{bmatrix}
         a\\
         \end{bmatrix},5\right),$$\\
$$b+c+a+a+f\left(\begin{bmatrix}
         b\\
         \end{bmatrix},5\right),$$\\
$$b+c+a+b+f\left(\begin{bmatrix}
         a\\
         \end{bmatrix},5\right),$$\\
$$b+c+b+a+f\left(\begin{bmatrix}
         a\\
         \end{bmatrix},5\right)$$\\  Since the level now is 5, we carry out the computation for only those characters that are present in row 5 of the "permitted matrix" ie a,b and c. Keeping in mind the prevention of the generation of repetitive permutations, the computation is:\\
$$a+a+b+b+c+f\left(\begin{bmatrix}
                    \\ 
                   \end{bmatrix},6\right),$$\\
$$a+a+b+c+b+f\left(\begin{bmatrix}
                    \\ 
                   \end{bmatrix},6\right),$$\\
$$a+b+a+b+c+f\left(\begin{bmatrix}
                    \\ 
                   \end{bmatrix},6\right),$$\\
$$a+b+a+c+b+f\left(\begin{bmatrix}
                    \\ 
                   \end{bmatrix},6\right),$$\\
$$a+b+b+a+c+f\left(\begin{bmatrix}
                    \\ 
                   \end{bmatrix},6\right),$$\\
$$a+b+b+c+a+f\left(\begin{bmatrix}
                    \\ 
                   \end{bmatrix},6\right),$$\\
$$a+c+a+b+b+f\left(\begin{bmatrix}
                    \\ 
                   \end{bmatrix},6\right),$$\\
$$a+c+b+a+b+f\left(\begin{bmatrix}
                    \\ 
                   \end{bmatrix},6\right),$$\\
$$a+c+b+b+a+f\left(\begin{bmatrix}
                    \\ 
                   \end{bmatrix},6\right),$$\\
$$b+a+a+b+c+f\left(\begin{bmatrix}
                    \\ 
                   \end{bmatrix},6\right),$$\\
$$b+a+a+c+b+f\left(\begin{bmatrix}
                    \\ 
                   \end{bmatrix},6\right),$$\\
$$b+a+b+a+c+f\left(\begin{bmatrix}
                    \\ 
                   \end{bmatrix},6\right),$$\\
$$b+a+b+c+a+f\left(\begin{bmatrix}
                    \\ 
                   \end{bmatrix},6\right),$$\\
$$b+b+a+a+c+f\left(\begin{bmatrix}
                    \\ 
                   \end{bmatrix},6\right),$$\\
$$b+b+a+c+a+f\left(\begin{bmatrix}
                    \\ 
                   \end{bmatrix},6\right),$$\\
$$b+c+a+a+b+f\left(\begin{bmatrix}
                    \\ 
                   \end{bmatrix},6\right),$$\\
$$b+c+a+b+a+f\left(\begin{bmatrix}
                    \\ 
                   \end{bmatrix},6\right),$$\\
$$b+c+b+a+a+f\left(\begin{bmatrix}
                    \\ 
                   \end{bmatrix},6\right)$$\\ The level is now 6 ie $n+1$. We have reached the base case of the recursive structure. We return a null character ie '$\backslash$0' (C++). Hence, the computation is:\\
$$aabbc,$$
$$aabcb,$$
$$ababc,$$
$$abacb,$$
$$abbac,$$
$$abbca,$$
$$acabb,$$
$$acbab,$$
$$acbba,$$
$$baabc,$$
$$baacb,$$
$$babac,$$
$$babca,$$
$$bbaac,$$
$$bbaca,$$
$$bcaab,$$
$$bcaba,$$
$$bcbaa$$\\
Which are the permutations of S under the given constraints.
\end{multicols}
\section{Procedure}
First, S is sorted in ascending order. This is done so that the permutations are generated in lexicographic order. (This step can be skipped if the user does not want the permutations to be generated in lexicographic order.) Now, a $N\times N$ matrix A is constructed such that:
\\
$A_{ij}=S{j}$
\\
Hence, A stores S in each of its rows.
\\
$$A=
  \begin{bmatrix}
  S_{1} & S_{2} & \cdots & S_{n} \\
  S_{2} & S_{2} & \cdots & S_{n} \\
  \vdots  & \vdots  & \ddots & \vdots  \\
  S_{1} & S_{2} & \cdots & S_{n}
 \end{bmatrix}$$
\\
The procedure is recursive in nature. The function resembles the determinant function. A modified version of the co-factor expansion is used, wherein a check against the "permitted matrix" is carried out and the expansion of only those characters that are present in the "permitted matrix" occurs. A new term called "level" is defined here, as follows:
\\$level=n+1 - $ (length of each row of the current matrix parameter of the function)\\
Level is another parameter to the function. 
\\
\\
Hence the computation would be of the form:
\\
\\
$$f\left(
  \begin{bmatrix}
  S_{1} & S_{2} & \cdots & S_{n} \\
  S_{2} & S_{2} & \cdots & S_{n} \\
  \vdots  & \vdots  & \ddots & \vdots  \\
  S_{1} & S_{2} & \cdots & S_{n}
 \end{bmatrix},1\right)$$
\\
$$\left(level=n+1-n=1\right)$$
\\
Now, to apply the constraints, a check is conducted against the (level)th row in the "permitted matrix." Suppose the permitted matrix is of the form:
\\
$$"permitted"= \begin{bmatrix}
                    S_{i} & S_{j} &    \\
                    S_{k} & S_{l} & S{m} \\
                    \vdots  & \vdots  & \ddots & \vdots  \\
                    \vdots  & \vdots  & \ddots & \vdots  \\
                    \end{bmatrix}
                    \\
                    \\
                     ; 1 \leq i,j,k,l,m \leq n$$
\\
\\
In the "permitted matrix", each row has distinct elements, which is quite obvious.
\\
\\
$\begin{bmatrix}
  S_{1} & S_{2} & \cdots & S_{n} \\
  \end{bmatrix}$
  is checked against $\begin{bmatrix}
                                 S_{i} & S_{j} &    \\
                                 \end{bmatrix}$\ and hence the co-factor expansion is carried out only for i and j.
\\
\\
Hence, the computation will be of the form
$$S_{i}+f\left(
        \begin{bmatrix}
        S_{1} & S_{2} & \cdots & S_{i-1} & S_{i+1} &\cdots & S_{n} \\
        S_{1} & S_{2} & \cdots & S_{i-1} & S_{i+1} &\cdots & S_{n} \\
        \vdots  & \vdots  & \ddots & \vdots  \\
        S_{1} & S_{2} & \cdots & S_{i-1} & S_{i+1} &\cdots & S_{n} \\
        \end{bmatrix},2\right)   ,   S_{j}+f\left(
        \begin{bmatrix}
        S_{1} & S_{2} & \cdots & S_{j-1} & S_{j+1} &\cdots & S_{n} \\
        S_{1} & S_{2} & \cdots & S_{j-1} & S_{j+1} &\cdots & S_{n} \\
        \vdots  & \vdots  & \ddots & \vdots  \\
        S_{1} & S_{2} & \cdots & S_{j-1} & S_{j+1} &\cdots & S_{n} \\
        \end{bmatrix},2\right)$$
        \\
$$\left(level=n+1-(n-1)=2\right)$$
Say S had repeated characters. These characters would always occur together as S is sorted. Consider $S_{a}=S_{b}=S_{i}$ ; $1 \leq a,b \leq n$. Yet, we write the computation as:
$$S_{i}+f\left(
        \begin{bmatrix}
        S_{1} & S_{2} & \cdots & S_{i-1} & S_{i+1} &\cdots & S_{n} \\
        S_{1} & S_{2} & \cdots & S_{i-1} & S_{i+1} &\cdots & S_{n} \\
        \vdots  & \vdots  & \ddots & \vdots  \\
        S_{1} & S_{2} & \cdots & S_{i-1} & S_{i+1} &\cdots & S_{n} \\
        \end{bmatrix},2\right)   ,   S_{j}+f\left(
        \begin{bmatrix}
        S_{1} & S_{2} & \cdots & S_{j-1} & S_{j+1} &\cdots & S_{n} \\
        S_{1} & S_{2} & \cdots & S_{j-1} & S_{j+1} &\cdots & S_{n} \\
        \vdots  & \vdots  & \ddots & \vdots  \\
        S_{1} & S_{2} & \cdots & S_{j-1} & S_{j+1} &\cdots & S_{n} \\
        \end{bmatrix},2\right)$$
\\
and not:
$$S_{i}+f\left(
        \begin{bmatrix}
        S_{1} & S_{2} & \cdots & S_{i-1} & S_{i+1} &\cdots & S_{n} \\
        S_{1} & S_{2} & \cdots & S_{i-1} & S_{i+1} &\cdots & S_{n} \\
        \vdots  & \vdots  & \ddots & \vdots  \\
        S_{1} & S_{2} & \cdots & S_{i-1} & S_{i+1} &\cdots & S_{n} \\
        \end{bmatrix},2\right),$$  \\ 
        \\
        $$S_{i}+f\left(
        \begin{bmatrix}
        S_{1} & S_{2} & \cdots & S_{i-1} & S_{i+1} &\cdots & S_{n} \\
        S_{1} & S_{2} & \cdots & S_{i-1} & S_{i+1} &\cdots & S_{n} \\
        \vdots  & \vdots  & \ddots & \vdots  \\
        S_{1} & S_{2} & \cdots & S_{i-1} & S_{i+1} &\cdots & S_{n} \\
        \end{bmatrix},2\right),$$   \\ 
        \\
        $$S_{j}+f\left(
        \begin{bmatrix}
        S_{1} & S_{2} & \cdots & S_{j-1} & S_{j+1} &\cdots & S_{n} \\
        S_{1} & S_{2} & \cdots & S_{j-1} & S_{j+1} &\cdots & S_{n} \\
        \vdots  & \vdots  & \ddots & \vdots  \\
        S_{1} & S_{2} & \cdots & S_{j-1} & S_{j+1} &\cdots & S_{n} \\
        \end{bmatrix},2\right)$$
\\
This ensures that repetitive permutations are not generated. This same point is implemented ahead too, at all levels, to ensure that repetitive permutations are not generated. 
\\
\\
Now, $\begin{bmatrix}
        S_{1} & S_{2} & \cdots & S_{i-1} & S_{i+1} &\cdots & S_{n} \\
       \end{bmatrix}$ is checked against the (level)th (ie 2nd) row in the "permitted matrix" $\begin{bmatrix}
                  S_{k} & S_{l} & S{m} & ...\\
                  \end{bmatrix}$ and hence the co-factor expansion is carried out only for $S_{k},S_{l}$ and $S_{m}$, in a manner similar to the above computation:
                  $$S_{i}+S_{k}+f\left(
        \begin{bmatrix}
        S_{1} & S_{2} & \cdots & S_{k-1} & S_{k+1} &\cdots & S_{n} \\
        S_{1} & S_{2} & \cdots & S_{k-1} & S_{k+1} &\cdots & S_{n} \\
        \vdots  & \vdots  & \ddots & \vdots  \\
        S_{1} & S_{2} & \cdots & S_{k-1} & S_{k+1} &\cdots & S_{n} \\
        \end{bmatrix},3\right),$$ \\
        \\
         $$S_{i}+S_{l}+f\left(
        \begin{bmatrix}
        S_{1} & S_{2} & \cdots & S_{l-1} & S_{l+1} &\cdots & S_{n} \\
        S_{1} & S_{2} & \cdots & S_{l-1} & S_{l+1} &\cdots & S_{n} \\
        \vdots  & \vdots  & \ddots & \vdots  \\
        S_{1} & S_{2} & \cdots & S_{l-1} & S_{l+1} &\cdots & S_{n} \\
        \end{bmatrix},3\right),$$ \\
        \\
         $$S_{i}+S_{m}+f\left(
        \begin{bmatrix}
        S_{1} & S_{2} & \cdots & S_{m-1} & S_{m+1} &\cdots & S_{n} \\
        S_{1} & S_{2} & \cdots & S_{m-1} & S_{m+1} &\cdots & S_{n} \\
        \vdots  & \vdots  & \ddots & \vdots  \\
        S_{1} & S_{2} & \cdots & S_{m-1} & S_{m+1} &\cdots & S_{n} \\
        \end{bmatrix},3\right),$$ \\
        \\
          $$S_{j}+S_{k}+f\left(
        \begin{bmatrix}
        S_{1} & S_{2} & \cdots & S_{k-1} & S_{k+1} &\cdots & S_{n} \\
        S_{1} & S_{2} & \cdots & S_{k-1} & S_{k+1} &\cdots & S_{n} \\
        \vdots  & \vdots  & \ddots & \vdots  \\
        S_{1} & S_{2} & \cdots & S_{k-1} & S_{k+1} &\cdots & S_{n} \\
        \end{bmatrix},3\right),$$ \\
        \\
         $$S_{j}+S_{l}+f\left(
        \begin{bmatrix}
        S_{1} & S_{2} & \cdots & S_{l-1} & S_{l+1} &\cdots & S_{n} \\
        S_{1} & S_{2} & \cdots & S_{l-1} & S_{l+1} &\cdots & S_{n} \\
        \vdots  & \vdots  & \ddots & \vdots  \\
        S_{1} & S_{2} & \cdots & S_{l-1} & S_{l+1} &\cdots & S_{n} \\
        \end{bmatrix},3\right),$$ \\
        \\
         $$S_{j}+S_{m}+f\left(
        \begin{bmatrix}
        S_{1} & S_{2} & \cdots & S_{m-1} & S_{m+1} &\cdots & S_{n} \\
        S_{1} & S_{2} & \cdots & S_{m-1} & S_{m+1} &\cdots & S_{n} \\
        \vdots  & \vdots  & \ddots & \vdots  \\
        S_{1} & S_{2} & \cdots & S_{m-1} & S_{m+1} &\cdots & S_{n} \\
        \end{bmatrix},3\right)$$ 
\\
$$\left(level=n+1-(n-2)=3\right)$$
\\
Hence, as the computation is carried out, a $1\times 1$ matrix is reached eventually. The "modified cofactor" of the element in the matrix is an empty matrix. Here the level is n+1.$\left(level=n+1-(0)=n+1\right)$ This is the base case of the recursive function. Null is returned (C++). Hence, the recursive structure is complete and the permutations are generated under the given restraints and in lexicographic order without repetitions.

\section{Worst-case analysis}
The worst case occurs when there are no constraints i.e. all permutations have to be generated. In this case, we can check if the forbidden matrix is empty. Then, it is equivalent to evaluating a determinant of size $n$ which follows the following recurrence relation:\\
$T(n)=nT(n-1)$ , $T(0)=1(=T(1))$\\
This is $\mathcal{O}(n!)$ in terms of time complexity.
\section{Conclusion}
Thus, a method to generate the permutations of a given string under a given set of constraints is established. This method is efficient because it does not generate all the permutations and then check which of them satisfy the given set of constraints. Instead, it directly generates only those permutations which satisfy the given set of constraints. Also, the permutations can be generated in lexicographic order by simply sorting the string before carrying out any computations. The repetitive permutations are not generated.\\

\end{document}